\documentclass[aps,prl,showpacs,preprintnumbers,
               amsmath,amssymb,11pt]{revtex4}
\usepackage{dcolumn}
\usepackage[colorlinks]{hyperref}

\newcommand{\be}{\begin{equation}}
\newcommand{\ee}{\end{equation}}
\newcommand{\ba}{\begin{array}}
\newcommand{\bqa}{\begin{eqnarray}}
\newcommand{\eqa}{\end{eqnarray}}
\newcommand{\Frac}[2]{\frac{\displaystyle #1}{\displaystyle #2}}

\newcommand{\cO}{{\cal O}}

\begin{document}
\title{\bf Resonance sum rules from large $N_C$ and partial wave dispersive analysis }

\author{Zhi-Hui~Guo }
\affiliation{  Department of Physics, Peking University, Beijing 100871, P.~R.~China  \\
and \\
 IFIC, ( CSIC-Universitat de Valencia ), Apt.~Correus 22085, E-46071, Valencia, Spain }

\begin{abstract}
Combining large $N_C$ techniques and partial wave dispersion theory to analyze the $\pi\pi$ scattering,
without relying on any explicit resonance lagrangian,  some interesting results are derived:
(a) a general KSRF relation including the scalar meson  contribution;
(b) a new relation between  resonance couplings, with which we have made an intensive analysis  in several specific models;
(c) low energy constants in chiral perturbation theory related with $\pi\pi$ scattering in terms of the mass and decay width of resonances.
\end{abstract}

\vskip .5cm


\date{\today}
\maketitle

Since QCD is non-perturbative in the low energy range, it is very hard to describe hadrons  from  first principle QCD.
Instead,  effective field theory and S-matrix techniques are widely used in this area. For effective field theory
the chiral symmetry can be incorporated , but it is not easy to incorporate a good high energy behavior.
In contrast, S-matrix theory can implement apparent high energy constraints, but to take into account the chiral symmetry is
not trivial. So the combination of  effective field theory and  S-matrix techniques are powerful and popular in hadron physics.
 Chiral Perturbation  Theory ($\chi$PT)  is a very successful  effective  theory of low energy QCD,
 which only contains the freedom of pseudo-Goldstone mesons \cite{chpt845}. However when the energy reaches around 1 GeV,
one has to include the  heavy resonance freedom explicitly, such a theory is called resonance chiral perturbation theory (R$\chi$PT) \cite{rchpt}.
On the S-matrix side, there have also been intensive studies, especially in $\pi\pi$ scattering process \cite{smatrix} .
In this note, we study the $\pi\pi$ scattering in the partial wave  dispersive theory,
combining with the large $N_C$ techniques \cite{lnc4}\cite{lnc6}.

Our starting point is the once-subtracted T-matrix dispersion relation,
\begin{eqnarray}
 T_J(s) -  T_J(0) = \frac{s}\pi\int_{-\infty}^0\frac{ds'
\mbox{Im}T_J(s')}{s'(s'-s)}\,  +\, \frac{s}{\pi}\int_{4
m_\pi^2}^\infty  \frac{ds' \, \mbox{Im}T_J(s')}{s'(s'-s)} \,,
\label{disponce}
\end{eqnarray}
where the first integral in the right hand side of the above equation is the left hand cut contribution ($t$-channel)
and the second integral corresponds to the right hand cut ($s$-channel).
In the large $N_C$ limit, the resonances become narrow-width states, which allows us to write down the imaginary part of the
$s$-channel resonance exchange,
\begin{equation} \label{imts}
\mbox{Im}T_J^{I, \rm R}(s)\, =\, \pi \,  \frac{M_{\rm R}\,
\Gamma_{\rm R} }{\rho_{\rm R} }\, \, \delta(s-M_{\rm R}^2)\,,
\end{equation}
where $\rho_{\rm R}=\sqrt{\frac{M_R^2-4m_\pi^2}{M_R^2}}$ and
the subscript $R$ denote the different resonances.

Crossing symmetry relates the right to the left-hand cut through the
expression~\cite{martin},
\begin{eqnarray}
 \label{imtl}
&& \hspace*{-0.5cm}
\mathrm{Im_L}T^{I}_{J}(s) =
\frac{1+(-1)^{I+J}}{s-4m_{\pi}^2}
\sum_{J'}\sum_{I'}(2J'+1)C^{st}_{II'} \times \nonumber \\ &&
\int_{4m_{\pi}^2}^{4m_{\pi}^2-s}dt
P_J(1+\frac{2t}{s-4m_{\pi}^2})  P_{J'}(1+\frac{2s}{t-4m_{\pi}^2})
\mathrm{Im_R}T^{I'}_{J'}(t)  ,
\end{eqnarray}
where $P_n(x)$ are the Legendre polynomials and the crossing matrix can be found in \cite{martin}.
Substituting Eq.(\ref{imts}) and Eq.(\ref{imtl}) into Eq.(\ref{disponce}), we can express the right hand side of
Eq.(\ref{disponce}) in terms of parameters of the resonances.

For the local value of  T-matrix, i.e. $T(s)$ and $T(0)$ in Eq.(\ref{disponce}), we use $\chi$PT results to estimate them,
 since later we will match the left and right hand sides of Eq.(\ref{disponce}) at low energy range.  Up to now,
we implement the high energy constraints through the once-subtracted dispersion relation and the chiral symmetry in
our method is reflected in the $\chi$PT results for $\pi\pi$ scattering amplitudes. Making the chiral expansion
on both sides of Eq.(\ref{disponce}) and matching them at different chiral orders, lead to the following relations:
\begin{itemize}
\item Matching at $O(p^2)$
\begin{eqnarray}\label{ksrf}
&&  \frac{144 \pi f^2 \overline{\Gamma}_V}{\overline{M}_V^3 }
\, +\, \frac{32\pi f^2 \overline{\Gamma}_S}{\overline{M}_S^3}\,
=\, 1\, ,
\end{eqnarray}
\item Matching at $O(p^4)$
\begin{eqnarray}
&& \frac{9 \overline{\Gamma}_V}{\overline{M}_V^5}
\left[\alpha_V+6\right]
\, +\, \Frac{2 \overline{\Gamma}_S}{3 \overline{M}_S^5 }
\left[ \alpha_S+6\right] \,
=\, 0\, ,   \nonumber
\nonumber \\ &&
 L_2 \, =\, 12 \pi f^4 \Frac{\overline{\Gamma}_V}{\overline{M}_V^5} ,\, \,\,\,\,
 L_3 \,=\, 4 \pi f^4
\left( \Frac{2\overline{\Gamma}_S}{3 \overline{M}_S^5} \,
-\,  \Frac{ 9 \overline{\Gamma}_V}{\overline{M}_V^5} \right) \, ,
\end{eqnarray}

\item Matching at $O(p^6)$
\bqa
 &&   r_2-2r_f=\frac{64\pi f^6 \overline{\Gamma}_S}{\overline{M}_S^7} \left( 1+
\frac{\beta_{\rm S}}{3} +\frac{\gamma_{\rm S}}{6}   \right)+\frac{\pi f^6
\overline{\Gamma}_V}{\overline{M}_V^7} \left( 7584 + 1248 \beta_{\rm
V} + 144 \gamma_{\rm V}       \right) \,, \nonumber \\&&
r_3=\frac{64 \pi f^6 \overline{\Gamma}_S}{3\overline{M}_S^7} \left(
1+\frac{\beta_{\rm S}}{2}\right) \, - \, \frac{768 \pi f^6
\overline{\Gamma}_V}{\overline{M}_V^7} ( 1 + \frac{3  \beta_{\rm
V}}{32} )  \,\,\, ,
 r_4=\frac{192 \pi f^6 \overline{\Gamma}_V}{\overline{M}_V^7}
 \left( 1 +\frac{\beta_{\rm V}}{8}\right) \,,
\nonumber \\&&
r_5=\frac{32 \pi f^6 \overline{\Gamma}_S}{3\overline{M}_S^7}+\frac{36 \pi f^6 \overline{\Gamma}_V}{\overline{M}_V^7}
\, ,
r_6=\frac{12 \pi f^6 \overline{\Gamma}_V}{\overline{M}_V^7} \,  .
\eqa
\end{itemize}
where  $L_i$ and $r_i$ are the $O(p^4)$ and $O(p^6)$ low energy constants in $\chi$PT respectively,
 see \cite{lnc6} and references therein for details; $\overline{\Gamma}_R$ and $\overline{M}_R$ stand, respectively,
for the value of the $R$ resonance width and mass in the chiral limit; the $O(m_\pi^2)$ corrections are reflected in $\alpha, \beta, \gamma$
\begin{equation}
\label{beta}
\frac{\Gamma_R}{M_R^5}=\frac{\overline{\Gamma}_R}{\overline{M}_R^{5}} \left[1+
\beta_R\frac{m_\pi^2}{\overline{M}_R^{2}}+\cO(m_\pi^4)\right], \ee \be
\label{alphagamma}
\frac{\Gamma_R}{M_R^3}=\frac{\overline{\Gamma}_R}{\overline{M}_R^{3}} \left[1+
\alpha_R\frac{m_\pi^2}{\overline{M}_R^{2}}+\gamma_R\frac{m_\pi^4}{\overline{M}_R^{4}}+\cO(m_\pi^6)\right].
\end{equation}

All the above relations are derived relying on both the proper high ( once-subtracted dispersion relation) and low ( $\chi$PT matching)
 energy behaviors and without relying on any explicit resonance lagrangian. Most of the effective lagrangians are constructed due to
 related symmetries, which means in this case the low energy behaviors are automatically satisfied. The once-subtracted
dispersion relation requires that the $\pi\pi$ scattering amplitudes should approach a constant or vanish in $s \rightarrow \infty$ ($s$ is the
invariant energy of $\pi\pi$ system ).  Some specific lagrangian models are analyzed
\begin{itemize}
\item Linear sigma model: In this case, the renormalizability ensures the proper high energy behavior and chiral invariant
ensures the low energy symmetry. All the above relations are automatically satisfied here.

\item Gauge chiral model: To recover the relations listed above, $\pi$-$a_1$ mixing must be taken into account.

\item R$\chi$PT and its extensions: Collecting the higher order operators into several effective couplings appearing
in the minimal R$\chi$PT, such as $G_V^{eff}, c_d^{eff}$, the $\pi\pi$ scattering amplitude at high energies is
\begin{equation}
T(s)\, = \,
\Frac{s}{96 \pi f_\pi^2}\,
\left[ 1\, -\,  \Frac{3 \, G_V^{{\rm eff} \, 2}}{f_\pi^2}
\, - \, \frac{2\, c_d^{{\rm eff}\, 2}}{f_\pi^2}\right]\,\,+\,\,\, \cO(s^0)\, .
\end{equation}
where we can see the coefficient of $s$ vanishes, since it is nothing but the general $KSRF$ relation Eq.(\ref{ksrf}) in this specific lagrangian.

\item Scalar tadpole contribution to $O(p^6)$ low energy constants:  there is an intensive study on the $O(p^6)$ low energy constants
\cite{rchtop6}. However our method provides another convenient procedure to derive the low energy constants related to $\pi\pi$ scattering.
Here we focus in the specific resonance lagrangian proposed in \cite{op61}. Using our method to calculate the $r_i$, we find that
the $r_i$ in \cite{op61} are derived without considering the scalar meson tadpole effects.
After taking into account the tadpole effects, the new values for $r_i$ are
\begin{equation}\label{newres}
r_2\, =\, 18 \cdot 10^{-4} \, , \quad r_3\,=\, 0.9 \cdot 10^{-4}\, ,
\quad r_4\, =\, -1.9\cdot 10^{-4}\, ,
\end{equation}
comparing with the values given in \cite{op61}
\begin{equation}\label{oldres}
r_2\, =\, 1.3\cdot 10^{-4} \, , \quad r_3\,=\, -1.7 \cdot 10^{-4}\,
, \quad r_4\, =\, -1.0\cdot 10^{-4}\, .
\end{equation}
Although there are large discrepancies in the $O(p^6)$ low energy constants, their effects on the global uncertainties in the scattering
length determinations are negligible.
\end{itemize}

We have presented a convenient procedure to implement the proper high and low  energy constraints for $\pi\pi$ scattering
and this method could be useful for different processes in future studies.
\section*{Acknowledgments}

I would like to thank the organizers of the workshop Scadron 70 "Scalar meson and related topics " for their kindness and the support of my residence.
I also would like to thank the Parsifal group in IFIC for partial support, their hospitality and
 especially for Jorge Portoles to revise the manuscript. Z.H.~Guo is supported by China Scholarship Council for the research stay in IFIC.
This work is also supported in part by the EU MRTN-CT-2006-035482 (FLAVIAnet),
by MEC (Spain) under grant FPA2007-60323 and by the Spanish Consolider-Ingenio 2010 Programme CPAN (CSD2007-00042).


\begin{thebibliography}{26}
\bibitem{chpt845} J. Gasser and H. Leutwyler, Annals Phys. {\bf 158} (1984) 142; J.~Gasser and H.~Leutwyler,
  {\it Nucl. Phys.} {\bf B  250} (1985) 465;
\bibitem{rchpt}G. Ecker {\it et al.}, Nucl. Phys. {\bf B321} (1989)311.
\bibitem{smatrix} Z.~G.~Xiao and H.~Q.~Zheng, Nucl. Phys. {\bf A 695} (2001) 273; H.~Q.~Zheng $et$ $al.$, Nucl. Phys. {\bf A 733 } (2004) 235;
Z.~Y.~Zhou, {\it et al. } JHEP 0502 (2005) 043; I.~Caprini, {\it et al.} Phys. Rev. Lett. 96 (2006) 132001.
\bibitem{lnc4} Z.~H.~Guo, J.~J.~Sanz~Cillero, H.~Q.~Zheng, JHEP 0706 (2007) 030.
\bibitem{lnc6} Z.~H.~Guo, J.~J.~Sanz~Cillero, H.~Q.~Zheng, arXiv: 0710.2163 [hep-ph], accetpted by Phys. Lett. B.
\bibitem{martin}B.~R.~Martin, D.~Morgan and G.~Shaw, {\it Pion Pion Interactions in
Particle Physics}, Academic Press, London, 1976
\bibitem{rchtop6} V. Cirigliano {\it et al.}, Nucl. Phys. {\bf B 753} (2006) 139.
\bibitem{op61} J.~Bijnens, {\it et al.}, Nucl.Phys.{\bf B 508}(1997) 263, Erratum-ibid.{\bf B 517}(1998) 639.


\end{thebibliography}
\end{document}